\documentclass[11pt]{article}

\usepackage{amssymb,amsmath}%,amsthm,amscd
\usepackage{epsf,epsfig}%,amsfonts..epsf,
\usepackage{texdraw}%pstricks,
\usepackage{color}

\newtheorem{Def}{Definition}[section]

\newtheorem{Rem}[Def]{Remark}

\newcommand{\br}{\begin{Rem}}
\newcommand{\er}{\end{Rem}}
\newtheorem{ex}[Def]{Example}
\newcommand{\bex}{\begin{ex}}
\newcommand{\eex}{\end{ex}}
\newcommand{\bd}{\begin{Def}}
\newcommand{\ed}{\end{Def}}

\newcommand{\be}{\begin{equation}}
\newcommand{\ee}{\end{equation}}
\newcommand{\bea}{\begin{eqnarray}}
\newcommand{\eea}{\end{eqnarray}}
\newcommand{\nn}{\nonumber}
\newcommand{\pa}{\partial}

\title{${\mathbb{Z}}_N$ graded discrete Lax pairs and Yang-Baxter maps}
\author{Allan P. Fordy\thanks{School of Mathematics,
University of Leeds, Leeds LS2 9JT, UK. ~~E-mail: a.p.fordy@leeds.ac.uk}
$\,$ and Pavlos Xenitidis\thanks{School of Mathematics,
Statistics and Actuarial Science, University of Kent, Canterbury CT2 7NF, UK. ~~E-mail: p.xenitidis@kent.ac.uk}
}

\date{\today}

\begin{document}

\maketitle

\begin{abstract}
We recently introduced a class of ${\mathbb{Z}}_N$ graded discrete Lax pairs and studied the associated discrete integrable systems (lattice equations). In this paper we introduce the corresponding Yang-Baxter maps. Many well known examples belong to this scheme for $N=2$, so,
for $N\geq 3$, our systems may be regarded as generalisations of these.

In particular, for each $N$ we introduce a generalisation of the map $H_{III}^B$ in the classification of scalar Yang-Baxter maps \cite{10-4}. For $N=3$ this is equivalent to the Yang-Baxter map associated with the discrete modified Boussinesq equation.  For $N\geq 5$ (and odd) we introduce a new family of Yang-Baxter maps, which have no lower dimensional analogue.  We also present multi-component versions of the Yang-Baxter maps $F_{IV}$ and $F_V$ (given in the classification of \cite{04-5}).
\end{abstract}


%\tableofcontents

%\bibliographystyle{plain}

\section{Introduction}

The term ``Yang-Baxter map'' was introduced by Veselov \cite{03-9} as an abbreviation for Drinfeld's notion of ``set-theoretical solutions to the quantum Yang–-Baxter equation''.  The basic ingredient is a map $R:\mathrm{X}\times \mathrm{X}\rightarrow \mathrm{X}\times \mathrm{X}$, where $\mathrm{X}$ is some algebraic variety.  For the case $\mathrm{X}= \mathbb{CP}^1$, these were partially classified in \cite{04-5,10-4}.  In \cite{06-6} a symmetry approach was introduced to relate Yang-Baxter equations with $3D$ consistent equations on quad-graphs, which had been classified in \cite{03-8}. Starting with any symmetry of an integrable equation on a quad-graph, the authors introduce invariant functions, which are then used to define a map. The Yang-Baxter relation was shown to be a {\em consequence} of $3D$ consistency.  Multicomponent Yang-Baxter maps are not yet classified, but several are known (see, for example, \cite{03-10,06-6,09-7,14-1}).

We recently introduced a class of ${\mathbb{Z}}_N$ graded discrete Lax pairs and studied the associated discrete integrable systems \cite{f14-3}.  Many well known examples belong to that scheme for $N=2$, so, for $N\geq 3$, our systems may be regarded as generalisations of these. As mentioned above, the quad systems for $N=2$ can be related to Yang-Baxter maps by the symmetry procedure of \cite{06-6}.  In this paper we construct generalisations of these, associated with our generalised lattice equations.

In Section \ref{basic} we present the basic background theory of Yang-Baxter maps and their relationship to lattice equations on a quadrilateral lattice.  In Section \ref{sec:ZN-LP}, we introduce the ${\mathbb{Z}}_N$-graded Lax pairs of \cite{f14-3} and derive the reduction to Yang-Baxter maps. We show that all such maps are equivalent to ones with ``level structure'' $(0,\delta;0,\delta)$.  These are studied in more detail in Section \ref{0d}, where some new families of Yang-Baxter maps are presented.  In Section \ref{addpot} we present a multicomponent generalisation of the Yang-Baxter maps $F_{IV}$ and $F_V$ (given in the classification of \cite{04-5})

\section{Basic Definitions}\label{basic}

Let $\mathrm{X}$ be an algebraic variety.  A {\em parametric Yang–-Baxter map} $R(a,b)$, depending upon parameters $(a,b)$, is a map
$$
R(a,b):\mathrm{X}\times \mathrm{X}\rightarrow \mathrm{X}\times \mathrm{X},
$$
satisfying:
\be\label{YBeq}
R_{23}(a_2,a_3)\circ R_{13}(a_1,a_3)\circ R_{12}(a_1,a_2) =
             {R}_{12}(a_1,a_2)\circ {R}_{13}(a_1,a_3)\circ {R}_{23}(a_2,a_3),
\ee
where ${R}_{ij}(a_i,a_j)$ is the map that acts as $R(a,b)$ on the $i$ and $j$ factor of $\mathrm{X}\times \mathrm{X}\times \mathrm{X}$, and identically on the other.

\bd[Reversibility]
Let $P$ be the involution given by $P({\bf x},{\bf y};a,b)=({\bf y},{\bf x};b,a)$.  If $P\circ R(a,b)$ is also an involution, then the map ${R}(a,b)$ is said to be {\em reversible}.  An alternative way of writing this is that the map $P\circ {R}(a,b)\circ P$ is the inverse of ${R}(a,b)$.
\ed

Lax pairs were defined for Yang-Baxter maps in \cite{03-9,03-10}.  A matrix $L({\bf x},a)$, with ${\bf x}\in \mathrm{X}$, depending upon the YB parameter $a$ and the spectral parameter $\lambda$ is used to define the equation:
\be\label{YBLax}
L({\bf x}',a)L({\bf y}',b) = L({\bf y},b)L({\bf x},a).
\ee
It was shown in \cite{03-9} that if $L$ satisfies this, then the map $({\bf x},{\bf y})\mapsto ({\bf x}',{\bf y}')$ satisfies the parametric Yang-Baxter equation (\ref{YBeq}) and is {\em reversible}.

\bd[The Companion Map]
The companion map $({\bf x},{\bf y}')\mapsto ({\bf x}',{\bf y})$ is obtained by solving equation (\ref{YBLax}) for the variables $({\bf x}',{\bf y})$.
\ed

\subsection{Reductions from a Lattice Equation}

Suppose we have a square lattice with vertices labelled $(m,n)$.  At each vertex we have functions
$$
{\bf u}_{m,n}=\left(u^{(0)}_{m,n},\dots,u^{(N-1)}_{m,n}\right),\quad {\bf v}_{m,n}=\left(v^{(0)}_{m,n},\dots,v^{(N-1)}_{m,n}\right),
$$
and vector function $\Psi_{m,n}$, satisfying
\be\label{LatticeLax}
\Psi_{m+1,n} = L({\bf u}_{m,n},a)\, \Psi_{m,n}, \quad \Psi_{m,n+1} = L({\bf v}_{m,n},b)\, \Psi_{m,n},
\ee
with compatibility conditions
\be\label{LatticeLax-eq}
L({\bf u}_{m,n+1},a) L({\bf v}_{m,n},b) =  L({\bf v}_{m+1,n},b) L({\bf u}_{m,n},a).
\ee
If we now consider the reduction
\be\label{reduce}
{\bf u}_{m,n} = {\bf x}_p,\quad {\bf v}_{m,n} = {\bf y}_{p+1},\quad\mbox{where}\quad p = n-m,
\ee
then (\ref{LatticeLax-eq}) reduces to (\ref{YBLax}), with ${\bf x}={\bf x}_p,\, {\bf x}'={\bf x}_{p+1},\, {\bf y}={\bf y}_p,\, {\bf y}'={\bf y}_{p+1}$, with the map $({\bf x},{\bf y})\mapsto ({\bf x}',{\bf y}')$ being Yang-Baxter.

\br
Notice that this does not rely on any underlying Lie point symmetry of the lattice equation.  It is just a ``travelling wave'' solution of the lattice equation.
\er

\section{${\mathbb{Z}}_N$-Graded Lax Pairs} \label{sec:ZN-LP}
\setcounter{equation}{0}

We now consider the specific discrete Lax pairs, which we introduced in \cite{f14-3}.  Consider a pair of matrix equations of the form
\begin{subequations}\label{eq:dLP-gen}
\bea
 \Psi_{m+1,n} &=& L_{m,n}\, \Psi_{m,n} \equiv \left( U_{m,n} + \lambda \,\Omega^{\ell_1}\right)\,\Psi_{m,n},  \label{eq:dLP-gen-L} \\
 \Psi_{m,n+1} &=& M_{m,n}\, \Psi_{m,n} \equiv \left( V_{m,n} + \lambda \,\Omega^{\ell_2}\right)\,\Psi_{m,n},  \label{eq:dLP-gen-M}
\eea
where
\be \label{eq:A-B-entries}
U_{m,n} = {\rm{diag}}\left(u^{(0)}_{m,n},\cdots,u^{(N-1)}_{m,n}\right) \Omega^{k_1},\quad
              V_{m,n} = {\rm{diag}}\left(v^{(0)}_{m,n},\cdots,v^{(N-1)}_{m,n}\right) \Omega^{k_2},
\ee
and
$$
(\Omega)_{i,j} \,=\, \delta_{j-i,1} +  \delta_{i-j,N-1}.
$$
\end{subequations}
The matrix $\Omega$ defines a grading and the four matrices of (\ref{eq:dLP-gen}) are said to be of respective levels $k_i, \ell_i$, with $\ell_i\neq k_i$ (for each $i$).  The Lax pair is characterised by the quadruple $\left(k_1,\ell_1;k_2,\ell_2 \right)$, which we refer to as {\em the level structure} of system, and for consistency, we require
\be\label{eq:dLP-nec-rel}
k_1 + \ell_2 \,\equiv\,k_2 + \ell_1 \; (\bmod N).
\ee
Since matrices $U$, $V$ and $\Omega$ are independent of $\lambda$, the compatibility condition of (\ref{eq:dLP-gen}),
\begin{equation} \label{eq:dLP-gen-cc}
L_{m,n+1} M_{m,n} \,=\,M_{m+1,n} L_{m,n},
\end{equation}
splits into the system
\begin{subequations} \label{eq:dLP-gen-scc}
\begin{eqnarray}
U_{m,n+1} V_{m,n} &=& V_{m+1,n} U_{m,n}\,, \label{eq:dLP-gen-scc-1}\\
U_{m,n+1} \Omega^{\ell_2} - \Omega^{\ell_2} U_{m,n}  &=& V_{m+1,n} \Omega^{\ell_1} - \Omega^{\ell_1} V_{m,n}, \label{eq:dLP-gen-scc-2}
\end{eqnarray}
\end{subequations}
which can be written explicitly as
\begin{subequations}  \label{eq:dLP-ex-cc}
\begin{eqnarray}
u^{(i)}_{m,n+1} v_{m,n}^{(i+k_1)} &=& v^{(i)}_{m+1,n} u^{(i+k_2)}_{m,n}\,,  \label{eq:dLP-ex-cc-1}\\
u^{(i)}_{m,n+1} - u_{m,n}^{(i+\ell_2)} &=& v^{(i)}_{m+1,n} - v^{(i+\ell_1)}_{m,n}\,,   \label{eq:dLP-ex-cc-2}
\end{eqnarray}
\end{subequations}
or, in a solved form, as
\begin{equation}  \label{eq:dLP-ex-cc-s}
u^{(i)}_{m,n+1} \,=\, \frac{u^{(i+\ell_2)}_{m,n} - v^{(i+\ell_1)}_{m,n}}{u^{(i+k_2)}_{m,n} - v^{(i+k_1)}_{m,n}}\, u^{(i+k_2)}_{m,n}\,, \quad
v^{(i)}_{m+1,n}\,=\, \frac{u^{(i+\ell_2)}_{m,n} - v^{(i+\ell_1)}_{m,n}}{u^{(i+k_2)}_{m,n} - v^{(i+k_1)}_{m,n}}\, v^{(i+k_1)}_{m,n},
\end{equation}
assuming that $u^{(i)}_{m,n} \ne v^{(j)}_{m,n}$ for all $i,j$.  In all the above formulae, $i,\, j$ are taken $(\bmod N)$.

It is easily seen that the quantities
\be\label{a-b}
a=\prod_{i=0}^{N-1} u_{m,n}^{(i)},\quad b = \prod_{i=0}^{N-1} v_{m,n}^{(i)} \quad\mbox{satisfy}\quad  \Delta_n(a)=\Delta_m(b) = 0,
\ee
where
$$
\Delta_m = {\cal{S}}_m-1,\quad \Delta_n = {\cal{S}}_n-1, \quad\mbox{with}\quad
     {\cal{S}}_m f_{m,n}=f_{m+1,n}, \quad {\cal{S}}_n f_{m,n}=f_{m,n+1}.
$$

\subsection{Reduction to Yang-Baxter Maps}

We can now employ the reduction (\ref{reduce}), using (\ref{a-b}) to replace the components $x^{(N-1)}_p,\, y^{(N-1)}_p$. This introduces parameters $a,\, b$ into the Lax matrices.  If we define
\be \label{XpYp}
X_p = {\rm{diag}}\left(x^{(0)}_p,\dots,x^{(N-1)}_p\right),\quad
              Y_{p} = {\rm{diag}}\left(y^{(0)}_{p},\dots,y^{(N-1)}_{p}\right),
\ee
then the compatibility condition (\ref{eq:dLP-gen-cc}) takes the form
\be\label{XpYpeq}
(X_{p+1}\Omega^{k_1}+\lambda \Omega^{\ell_1})(Y_{p+1}\Omega^{k_2}+\lambda \Omega^{\ell_2}) =
       (Y_p\Omega^{k_2}+\lambda \Omega^{\ell_2})(X_p\Omega^{k_1}+\lambda \Omega^{\ell_1}),
\ee
and equations (\ref{eq:dLP-ex-cc}) take the form
\be\label{xpypeqs}
x^{(i)}_{p+1} y^{(i+k_1)}_{p+1} = y^{(i)}_{p} x^{(i+k_2)}_{p},\quad x^{(i)}_{p+1} + y^{(i+\ell_1)}_{p+1} = y^{(i)}_{p} + x^{(i+\ell_2)}_{p}.
\ee
We can write (\ref{XpYpeq}) as
\be\label{XpYp0d}
(X_{p+1}+\lambda \Omega^{\delta})(\Omega^{k_1}Y_{p+1}\Omega^{-k_1}+\lambda \Omega^{\delta}) =
       (Y_p+\lambda \Omega^{\delta})(\Omega^{k_2}X_p\Omega^{-k_2}+\lambda \Omega^{\delta}),
\ee
where $0<\delta \leq N-1$, with $\delta \equiv \ell_i-k_i \;(\bmod N)$.  This allows us to reduce the general case with level structure $(k_1,\ell_1;k_2,\ell_2)$ to that with level structure $(0,\delta;0,\delta)$.  First, note that formula (\ref{XpYp0d}) can be written
\be\label{XpYp0dbar}
(\bar X_{p+1}+\lambda \Omega^{\delta})(\bar Y_{p+1}+\lambda \Omega^{\delta}) =
       (\bar Y_p+\lambda \Omega^{\delta})(\bar X_p+\lambda \Omega^{\delta}),
\ee
where
$$
\bar X_p = {\rm{diag}}\left(\bar x^{(0)}_p,\dots,\bar x^{(N-1)}_p\right),\quad
              \bar Y_{p} = {\rm{diag}}\left(\bar y^{(0)}_{p},\dots,\bar y^{(N-1)}_{p}\right).
$$
Comparing (\ref{XpYp0dbar}) and (\ref{XpYp0d}), we see that
$$
\bar x^{(i)}_{p+1} =  x^{(i)}_{p+1}, \quad \bar y^{(i)}_{p+1} = y^{(i+k_1)}_{p+1}, \quad
    \bar x^{(i)}_{p} =  x^{(i+k_2)}_{p}, \quad \bar y^{(i)}_{p} = y^{(i)}_{p},
$$
all taken $(\bmod N)$.   We see from (\ref{XpYp0dbar}) that the components $(\bar x^{(i)}_{p},\bar y^{(i)}_{p})$ satisfy
$$
\bar x^{(i)}_{p+1} \bar y^{(i)}_{p+1} = \bar y^{(i)}_{p} \bar x^{(i)}_{p},\quad
               \bar x^{(i)}_{p+1} + \bar y^{(i+\delta)}_{p+1} = \bar y^{(i)}_{p} + \bar x^{(i+\delta)}_{p},
$$
which are just (\ref{xpypeqs}) with $(k_i,\ell_i)=(0,\delta)$.

\section{The Yang-Baxter Map Corresponding to the Case $(0,\delta ;0,\delta)$}\label{0d}
\setcounter{equation}{0}

In this section we consider the Lax equations with level structure $(0,\delta ;0,\delta)$, with $0<\delta \leq N-1$.  The resulting equations are {\em quadrirational}, with both the Yang-Baxter and companion maps being {\em birational}.  We find that the Yang-Baxter maps corresponding to $\delta$ and $N-\delta$ are inverses to each other and that the companion map is periodic, with period $N$.

\subsection{The Equations and Maps}

With Lax matrices
\be\label{Lax-0d}
L({\bf x},a) = X_p + \lambda \Omega^\delta,\quad L({\bf y},b) = Y_{p} + \lambda \Omega^\delta,
\ee
where $X_p$ and $Y_{p}$ are defined by (\ref{XpYp}), with
\be\label{xyn-1}
x_p^{(N-1)} = \frac{a}{\prod_{i=0}^{N-2} x_p^{(i)}},\quad y_{p}^{(N-1)} = \frac{b}{\prod_{i=0}^{N-2} y_{p}^{(i)}},
\ee
the Lax equation (\ref{YBLax}) implies
\be\label{xpyp0deqs}
x^{(i)}_{p+1} y^{(i)}_{p+1} = y^{(i)}_{p} x^{(i)}_{p},\quad x^{(i)}_{p+1} + y^{(i+\delta)}_{p+1} = y^{(i)}_{p} + x^{(i+\delta)}_{p}, \quad 0\leq i \leq N-1.
\ee
Only the formulae with $0\leq i\leq N-2$ are independent, but the full set is useful when discussing first integrals.

\br[Level structure $(\delta,0 ;\delta,0)$ vs $(0,\delta ;0,\delta)$]
Under the point transformation
$$
x^{(i)}_{p+1}=\tilde x^{(i+\delta)}_p,\quad x^{(i)}_{p}=\tilde x^{(i)}_{p+1},\quad
     y^{(i)}_{p+1}=\tilde y^{(i)}_p,\quad y^{(i)}_{p}=\tilde y^{(i+\delta)}_{p+1},
$$
equations (\ref{xpyp0deqs}) take the form
$$
\tilde x^{(i)}_{p+1} \tilde y^{(i+\delta)}_{p+1} = \tilde y^{(i)}_{p} \tilde x^{(i+\delta)}_{p},\quad \tilde x^{(i)}_{p+1} + \tilde y^{(i)}_{p+1} = \tilde y^{(i)}_{p} + \tilde x^{(i)}_{p}, \quad 0\leq i \leq N-1,
$$
which are just the equations for level structure $(\delta,0 ;\delta,0)$, so these structures are equivalent.
\er

\subsubsection{The Yang-Baxter map $R^{(\delta)}(a,b)$}

Here we solve (\ref{xpyp0deqs}) for $(x^{(i)}_{p+1}, y^{(i)}_{p+1})$ as functions of $(x^{(i)}_{p}, y^{(i)}_{p})$ (with $0\leq i\leq N-2$ and $x_p^{(N-1)},\, y_p^{(N-1)}$ replaced by (\ref{xyn-1})).  We write this map as $R^{(\delta)}(a,b)$, but when no ambiguity can arise, we suppress the parametric dependence by writing the map as $R^{(\delta)}$.

Notice that by shifting $i\mapsto i+N-\delta\equiv i-\delta\;(\bmod N)$, the second part of equation (\ref{xpyp0deqs}) takes the form
$$
x^{(i-\delta)}_{p+1} + y^{(i)}_{p+1} = y^{(i-\delta)}_{p} + x^{(i)}_{p},
$$
which means that the Yang-Baxter map $R^{(-\delta)}(a,b)$ is just the \underline{inverse} of $R^{(\delta)}(a,b)$.

\br[Non-Coprime Case]\label{ncp-remark}
When $\delta$ is a divisor of $N$, the map decouples into lower dimensional ones.
\er

This Yang-Baxter map has the following $N$ first integrals:
\be\label{ybfirst-int}
x^{(i)}_{p} y^{(i)}_{p} = c_i,\;\;\; 0\leq i \leq N-2,\quad \sum_{i=0}^{N-1} (x^{(i+\delta)}_{p} + y^{(i)}_{p}) = c_{N-1},
\ee
where, in the latter, $x_p^{(N-1)}$ and $y_p^{(N-1)}$ are replaced by (\ref{xyn-1}).  The last of these integrals is obtained by summing the additive equations of (\ref{xpyp0deqs}).

\subsubsection{The Companion Map $\varphi^{(\delta)}$}

Here we solve (\ref{xpyp0deqs}) for $(x^{(i)}_{p+1}, y^{(i)}_{p})$ as functions of $(x^{(i)}_{p}, y^{(i)}_{p+1})$ (with $0\leq i\leq N-2$ and $x_p^{(N-1)},\, y_p^{(N-1)}$ replaced by (\ref{xyn-1})).  Since $p$ is no longer the evolution parameter, we relabel our variables as:
$$
(x^{(i)}_{p}, y^{(i)}_{p+1})=(x^{(i)}_{q}, y^{(i)}_{q}), \quad (x^{(i)}_{p+1}, y^{(i)}_{p})=(x^{(i)}_{q+1}, y^{(i)}_{q+1}).
$$

\br[A second travelling wave reduction]
This labelling follows directly from the travelling wave reduction
$$
{\bf u}_{m,n} = {\bf x}_q,\quad {\bf v}_{m,n} = {\bf y}_{q},\quad\mbox{where}\quad q = n+m
$$
\er

We can re-arrange the quadratic formulae in (\ref{xpyp0deqs}) (with this new labelling) to obtain $N-1$ first integrals:
\be\label{comp-first-int1}
\frac{x^{(i)}_{q}}{y^{(i)}_{q}}=c_i,\;\;\; 0\leq i \leq N-2 .
\ee
We can also re-arrange the linear formulae of (\ref{xpyp0deqs}) to obtain
$$
x^{(i)}_{q+1}-y^{(i)}_{q+1} = x^{(i+\delta)}_{q} - y^{(i+\delta)}_{q}, \;\;\; 0\leq i \leq N-1 .
$$
If we define
\be\label{fxy}
f(x,y) = x-y,
\ee
then
\be\label{period}
f(x^{(i)}_{q+1},y^{(i)}_{q+1}) = f(x^{(i+\delta)}_{q},y^{(i+\delta)}_{q}),\quad 0\leq i\leq N-1.
\ee
We may use
$$
\left(\frac{x^{(0)}_{q}}{y^{(0)}_{q}},\dots ,\frac{x^{(N-2)}_{q}}{y^{(N-2)}_{q}},f\left(x^{(0)}_{q},y^{(0)}_{q}\right),\dots , f\left(x^{(N-2)}_{q},y^{(N-2)}_{q} \right)\right)
$$
as coordinates and, in these coordinates, the map $\varphi^{(\delta)}$ just shifts the coordinates $f(x^{(i)}_{q},y^{(i)}_{q})$ by $\delta$, whilst leaving the coordinates $\frac{x^{(i)}_{q}}{y^{(i)}_{q}}$ fixed.  With $(N,\delta)=1$ (see Remark \ref{ncp-remark}) this means that the map $\varphi^{(\delta)}$ is periodic with period $N$.
Furthermore, we have that $\varphi^{(\delta)}=\varphi^{(1)}\circ\cdots\circ\varphi^{(1)}$ (the $\delta-$fold composition of $\varphi^{(1)}$).  This statement is, of course, independent of coordinates.

\br[$(2N-2)$ first integrals]
Any cyclically symmetric function of $f(x^{(i)}_{q},y^{(i)}_{q})$ is a first integral of the companion map, so it possesses $(2N-2)$ first integrals.  The common level set is then finite, corresponding to the periodicity of the map.
\er

\subsection{Examples}

We can build hierarchies of Yang-Baxter maps for each $\delta$.

\subsubsection{The Case $N=2$}

Here we only have the case $\delta=1$, which leads to (with $x^{(0)}=x,\, y^{(0)}=y,\, x^{(1)}=a/x,\, y^{(1)}=b/y$)
$$
x_{p+1}=y_p \left(\frac{a+xy}{b+xy}\right),\quad y_{p+1}=x_p \left(\frac{b+xy}{a+xy}\right),
$$
which (up to a relabelling of parameters) is just the map $H_{III}^B$ in the classification of scalar Yang-Baxter maps \cite{10-4}.

The existence of the two invariant functions (\ref{ybfirst-int}) implies (the well known fact) that this map is an involution.

\br[The case $\delta=1$]
The system corresponding to $\delta=1$ exists for all $N\geq 2$, which can therefore be considered as a multi-component generalisation of the scalar Yang-Baxter map $H_{III}^B$.
\er

\subsubsection{The Case $N=3$}

Here we have $\delta=1$ and $\delta=2$, but since $N-1=2\equiv -1\, (\bmod 3)$, the map $R^{(2)}$ is just the inverse of $R^{(1)}$.  In this case $R^{(1)}$ takes the form:
\be\label{r1-3}
x^{(i)}_{p+1} = y^{(i)}_p \frac{A^{(i)}}{A^{(i+1)}},\qquad y^{(i)}_{p+1} = x^{(i)}_p \frac{A^{(i+1)}}{A^{(i)}},\;\;\; 0\leq i\leq 1,
\ee
with upper indices taken  $(\bmod 2)$ and where
$$
A^{(0)} = a(x^{(1)}_p+y^{(0)}_p)+x^{(0)}_p x^{(1)}_p y^{(0)}_p y^{(1)}_p,\quad A^{(1)} = A^{(0)}+(b-a) x^{(1)}_p ,\quad
        A^{(2)} = A^{(1)}+(b-a) y^{(0)}_p.
$$

\br[Discrete Modified Boussinesq Equation]
This is equivalent to the Yang-Baxter map derived in \cite{06-6}, associated with the discrete modified Boussinesq equation (see equation (67a-b) of \cite{06-6}).  They are related by a simple point transformation:
$$
x^{(0)}\mapsto \frac{c_0}{x^{1}},\quad x^{(1)}\mapsto c_0 x^{2}, \quad
              y^{(0)}\mapsto \frac{c_0\alpha_1}{\alpha_2 y^{1}},\quad y^{(1)}\mapsto \frac{\alpha_1^2 y^{2}}{c_0^3},\quad\mbox{where}\quad c_0^4=\frac{\alpha_1^3}{\alpha_2}.
$$
\er

\subsubsection{The Case $N=5$}

Here $\delta=1$ and $\delta=2$ give genuinely different maps.

The map $R^{(1)}$ takes the form:
\be\label{r1-5}
x^{(i)}_{p+1} = y^{(i)}_p \frac{A^{(i)}}{A^{(i+1)}},\qquad y^{(i)}_{p+1} = x^{(i)}_p \frac{A^{(i+1)}}{A^{(i)}},\;\;\; 0\leq i\leq 3,
\ee
with upper indices taken  $(\bmod 4)$ and where
\bea
&& A^{(0)} = a(x^{(1)}_p x^{(2)}_p x^{(3)}_p+y^{(0)}_p x^{(2)}_p x^{(3)}_p+y^{(0)}_p y^{(1)}_p x^{(3)}_p+y^{(0)}_p y^{(1)}_p y^{(2)}_p)
               + \prod_{i=0}^3 x^{(i)}_p y^{(i)}_p ,\nn\\
&&  A^{(1)}= A^{(0)}+(b-a) x^{(1)}_p x^{(2)}_p x^{(3)}_p,\;\;\; A^{(2)}= A^{(1)}+(b-a) y^{(0)}_p x^{(2)}_p x^{(3)}_p,\nn\\
 &&     A^{(3)}=A^{(2)} + (b-a) y^{(0)}_p y^{(1)}_p x^{(3)}_p,\;\;\; A^{(4)}=A^{(3)} + (b-a) y^{(0)}_p y^{(1)}_p y^{(2)}_p.  \nn
\eea

The map $R^{(2)}$ takes the form:
\be\label{r2-5}
x^{(0)}_{p+1} = y^{(0)}_p \frac{A^{(2)}}{A^{(3)}},\quad x^{(1)}_{p+1} = y^{(1)}_p \frac{A^{(0)}}{A^{(1)}},\quad
    x^{(2)}_{p+1} = y^{(2)}_p \frac{A^{(3)}}{A^{(4)}},\quad x^{(3)}_{p+1} = y^{(3)}_p \frac{A^{(1)}}{A^{(2)}},
\ee
and $y^{(i)}_{p+1} = \frac{x^{(i)}_p y^{(i)}_p}{x^{(i)}_{p+1}}$, with upper indices taken  $(\bmod 4)$ and where
\bea
&& A^{(0)} = a(x^{(0)}_p x^{(2)}_p x^{(3)}_p+x^{(0)}_p x^{(2)}_p y^{(1)}_p+x^{(2)}_p y^{(1)}_p y^{(3)}_p+y^{(0)}_p y^{(1)}_p y^{(3)}_p)
               + \prod_{i=0}^3 x^{(i)}_p y^{(i)}_p ,\nn\\
&&  A^{(1)}= A^{(0)}+(b-a) x^{(0)}_p x^{(2)}_p x^{(3)}_p,\;\;\; A^{(2)}= A^{(1)}+(b-a) x^{(0)}_p x^{(2)}_p y^{(1)}_p,\nn\\
 &&     A^{(3)}=A^{(2)} + (b-a) x^{(2)}_p y^{(1)}_p y^{(3)}_p,\;\;\; A^{(4)}=A^{(3)} + (b-a) y^{(0)}_p y^{(1)}_p y^{(3)}_p.  \nn
\eea

\br
Each $\delta$ introduces a new family of Yang-Baxter maps.  The case $\delta=3$ exists for all $N\geq 7$ and not a multiple of $3$.
\er

\subsection{The Quotient Potential Case and Symmetries}

In \cite{f14-3} we introduced two potential forms of our equations (\ref{eq:dLP-ex-cc}).  Here we briefly mention the ``quotient potential'', leaving the ``additive potential'' to Section \ref{addpot}.

Equations (\ref{eq:dLP-ex-cc-1}) hold identically if we set
\begin{equation} \label{eq:dLP-gen-ch-1}
u^{(i)}_{m,n}\,= \, \alpha\,\frac{\phi^{(i)}_{m+1,n}}{\phi^{(i+k_1)}_{m,n}}\,,\quad
            v^{(i)}_{m,n}\,= \, \beta\,\frac{\phi^{(i)}_{m,n+1}}{\phi^{(i+k_2)}_{m,n}},
\end{equation}
where $a=\alpha^N,\, b = \beta^N$.  Equations (\ref{eq:dLP-ex-cc-2}) then take the form
\begin{equation} \label{eq:dLP-gen-sys-1}
\alpha \left(\frac{\phi^{(i)}_{m+1,n+1}}{\phi^{(i+k_1)}_{m,n+1}}\,-\,\frac{\phi^{(i+\ell_2)}_{m+1,n}}{\phi^{(i+\ell_2+k_1)}_{m,n}} \right) \,=\,
  \beta \left(\frac{\phi^{(i)}_{m+1,n+1}}{\phi^{(i+k_2)}_{m+1,n}}\,-\,\frac{\phi^{(i+\ell_1)}_{m,n+1}}{\phi^{(i+\ell_1+k_2)}_{m,n}} \right),
\end{equation}
where indices are taken $(\bmod N)$.

These equations have a weighted scaling symmetry, whose invariants are given exactly by the formulae (\ref{eq:dLP-gen-ch-1}), leading us back to equations (\ref{eq:dLP-ex-cc}) and therefore to our previous Yang-Baxter maps.

\section{The Additive Potential}  \label{addpot}
\setcounter{equation}{0}

Equations (\ref{eq:dLP-ex-cc-2}) hold identically if we set
\be\label{eq:dLP-gen-ch-2}
u^{(i)}_{m,n} = \chi_{m+1,n}^{(i)}-\chi_{m,n}^{(i+\ell_1)},\quad
v^{(i)}_{m,n} = \chi_{m,n+1}^{(i)}-\chi_{m,n}^{(i+\ell_2)}.
\ee
Equations (\ref{eq:dLP-ex-cc-1}) then take the form
\be\label{eq:dLP-gen-sys-2}
\frac{\left( \chi^{(i)}_{m+1,n+1} - \chi^{(i+\ell_1)}_{m,n+1}\right)}{\left( \chi^{(i)}_{m+1,n+1} - \chi^{(i+\ell_2)}_{m+1,n}\right)}=
    \frac{\left( \chi^{(i+k_2)}_{m+1,n} - \chi^{(i+k_2+ \ell_1)}_{m,n}\right)}{\left( \chi^{(i+k_1)}_{m,n+1} - \chi^{(i+k_1+ \ell_2)}_{m,n}\right)} ,
\ee
and the first integrals (\ref{a-b}) take the form
\be\label{eq:dLP-gen-ch-2-cd}
\prod_{i=0}^{N-1} \left(\chi^{(i)}_{m+1,n} - \chi^{(i+\ell_1)}_{m,n} \right)\,=\,a,\quad
\prod_{i=0}^{N-1} \left(\chi^{(i)}_{m,n+1} - \chi^{(i+\ell_2)}_{m,n} \right)\,=\,b.
\ee
\br[Reduction]
It is not always possible to use these first integrals to explicitly reduce (\ref{eq:dLP-gen-sys-2}) to a system with $N-1$ components (eliminating $\chi^{(N-1)}_{m,n}$), and even when this is possible the spectral problem (\ref{eq:dLP-gen}) cannot be written in terms of the reduced variables.
\er
In \cite{f14-3} we showed that it is possible to explicitly reduce the system with $(k_i,\ell_i)=(0,1)$, which takes the form
\begin{subequations}\label{eq:h-0101-a}
\bea
 && \frac{\left( \chi^{(i)}_{m+1,n+1} - \chi^{(i+1)}_{m,n+1}\right)}{\left( \chi^{(i)}_{m+1,n+1} - \chi^{(i+1)}_{m+1,n}\right)} =
    \frac{\left( \chi^{(i)}_{m+1,n} - \chi^{(i+ 1)}_{m,n}\right)}{\left( \chi^{(i)}_{m,n+1} - \chi^{(i+ 1)}_{m,n}\right)} ,
                 \quad i = 0,\dots,N-3,\label{eq:h-0101-a1}  \\[3mm]
&&  \chi^{(N-2)}_{m+1,n+1} = \chi^{(0)}_{m,n}\,+\,\frac{1}{\chi_{m+1,n}^{(N-2)} - \chi_{m,n+1}^{(N-2)}}\,\left(\frac{a}{X}\,
                        -\,\frac{b}{Y} \right),  \label{eq:h-0101-a2}
\eea
\end{subequations}
where $X=\prod_{j=0}^{N-3}(\chi_{m+1,n}^{(j)}-\chi^{(j+1)}_{m,n})$ and $Y=\prod_{j=0}^{N-3}(\chi_{m,n+1}^{(j)}-\chi^{(j+1)}_{m,n})$.

\br
This is a direct generalisation of equation $H1$ in the ABS classification \cite{03-8}.
\er

It is easy to see that the system (\ref{eq:h-0101-a}) has the following pair of symmetry generators:
\begin{subequations}\label{Xts}
\bea
 {\bf X}_t &=& \sum_{i=0}^{N-2} \omega^{m+n+i} \pa_{\chi^{(i)}_{m,n}},   \label{Xt}  \\[3mm]
{\bf X}_s &=& \sum_{i=0}^{N-2} \omega^{m+n+i} \chi_{m,n}^{(i)} \pa_{\chi^{(i)}_{m,n}}, \;\; \omega\neq 1,  \label{Xs}
\eea
\end{subequations}
where $\omega^N=1$.  It is therefore possible to write equations (\ref{eq:h-0101-a}) in terms of the invariants of these symmetries.  We can then reduce this form of the lattice equations to Yang-Baxter maps.

\subsection{The Invariants of ${\bf X}_t$}

\begin{figure}[hbt]
\begin{center}
\unitlength=0.5mm
\begin{picture}(60,60)
%
%===========quad1=================
%
\put(0,0){\circle*{4}}
\put(50,0){\circle*{4}}
\put(0,50){\circle*{4}}
\put(50,50){\circle*{4}}
\thicklines
%\put(0,0){\vector(1,0){12}}
\put(0,0){\line(1,0){50}}
%\put(0,0){\vector(0,1){12}}
\put(0,50){\line(1,0){50}}
%\put(0,20){\vector(1,0){12}}
\put(50,0){\line(0,1){50}}
%\put(20,0){\vector(0,1){12}}
\put(0,0){\line(0,1){50}}
\put(25,-5){\makebox(0,0){$\bf x$}}
\put(-5,25){\makebox(0,0){$\bf v$}}
\put(25,55){\makebox(0,0){$\bf u$}}
\put(55,25){\makebox(0,0){$\bf y$}}
\put(-5,-5){\makebox(0,0){$\chi_{m,n}$}}
\put(57,-5){\makebox(0,0){$\chi_{m+1,n}$}}
\put(-5,55){\makebox(0,0){$\chi_{m,n+1}$}}
\put(60,55){\makebox(0,0){$\chi_{m+1,n+1}$}}
\end{picture}
\end{center}
\caption{Invariants defined on edges}\label{xyuv-fig}
\end{figure}
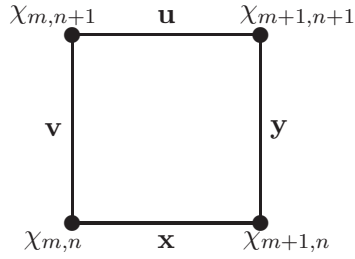

It is straightforward to write a suitable ``basis'' for the invariants of ${\bf X}_t$.  The formulae are more symmetric if we write ``too many'' invariants, which then satisfy some additional identities.  We therefore define $4(N-1)$ invariants, satisfying $(N-1)$ identities.  Furthermore, we make the reduction (\ref{reduce}), so that we derive a map. Following \cite{06-6}, we denote these invariants by
\be\label{xip}
x^{(i)} \equiv x^{(i)}_p,\;\; y^{(i)} \equiv y^{(i)}_p,\;\;
            u^{(i)} \equiv x^{(i)}_{p+1},\;\; v^{(i)} \equiv y^{(i)}_{p+1},\quad\mbox{where}\;\; p=n-m,
\ee
corresponding to specific edges of the lattice square, as shown in Figure \ref{xyuv-fig} and noting that the shifts $m\mapsto m-1$ and $n\mapsto n+1$ both correspond to $p\mapsto p+1$.

The $4(N-1)$ invariants:
\bea
&&  x^{(i)} = \chi^{(i)}_{m+1,n} - \chi^{(i+1)}_{m,n},\qquad i = 0,\dots,N-3, \quad
    x^{(N-2)}  =  \chi^{(N-2)}_{m+1,n} + \sum_{j=0}^{N-2} \chi_{m,n}^{(j)}, \nn \\[-1mm]
&&  y^{(i)} = \chi^{(i)}_{m+1,n+1} - \chi^{(i+1)}_{m+1,n},\quad i = 0,\dots,N-3,  \quad
    y^{(N-2)}  =  \chi^{(N-2)}_{m+1,n+1} + \sum_{j=0}^{N-2} \chi_{m+1,n}^{(j)}, \nn\\[-1mm]
&&  u^{(i)} = \chi^{(i)}_{m+1,n+1} - \chi^{(i+1)}_{m,n+1},\quad i = 0,\dots,N-3, \quad
    u^{(N-2)}  =  \chi^{(N-2)}_{m+1,n+1} + \sum_{j=0}^{N-2} \chi_{m,n+1}^{(j)}, \nn \\[-1mm]
&&  v^{(i)} = \chi^{(i)}_{m,n+1} - \chi^{(i+1)}_{m,n},\qquad i = 0,\dots,N-3,  \quad
    v^{(N-2)}  =  \chi^{(N-2)}_{m,n+1} + \sum_{j=0}^{N-2} \chi_{m,n}^{(j)}, \nn
\eea
satisfy $(N-1)$ identities:
\begin{subequations}\label{xyuvt-eq}
\bea
  x^{(i+1)}+ y^{(i)} &=& u^{(i)} + v^{(i+1)} ,\quad i = 0,\dots,N-3, \label{xyuvt-eq1}\\
    y^{(N-2)} + \sum_{j=0}^{N-2}v^{(j)} &=& u^{(N-2)} + \sum_{j=0}^{N-2} x^{(j)},  \label{xyuvt-eq2}
\eea
and equations (\ref{eq:h-0101-a}) take the form
\bea
  u^{(i)}v^{(i)} &=& x^{(i)} y^{(i)} ,\quad i = 0,\dots,N-3, \label{xyuvt-eq3}\\
    u^{(N-2)} &=& \sum_{j=0}^{N-2} v^{(j)} + \frac{1}{x^{(N-2)}-v^{(N-2)}} \left(\frac{a}{\prod_{j=0}^{N-3} x^{(j)}}
                   - \frac{b}{\prod_{j=0}^{N-3} v^{(j)}} \right).  \label{xyuvt-eq4}
\eea
\end{subequations}
The Yang-Baxter map corresponds to the solution of equations (\ref{xyuvt-eq}) for $(u^{(i)},v^{(i)})$.  We do not have an explicit form of the solution in general, but for any given value of $N$, this can be found.
\br[The Case $N=2$]
We already remarked that for $N=2$ the lattice equation is just $H1$ in the ABS classification \cite{03-8}.  Using the symmetry ${\bf X}_t$, with $\omega = -1$ leads to the Yang-Baxter map
$$
u = y + \frac{a-b}{x-y}, \quad v = x + \frac{a-b}{x-y},
$$
which is just $F_V$ of the ABS classification of quadrirational maps \cite{04-5} (the Adler map).  Clearly, we may consider this whole family of maps as multi-component generalisations of $F_V$.
\er
\bex[The Case $N=3$]  {\em
In this case, we find
\bea
 u^{(0)} &=& y^{(0)} + \frac{(a - b) y^{(0)}}{b - x^{(0)} y^{(0)} (x^{(0)} + x^{(1)} - y^{(1)})}, \nn \\
  u^{(1)} &=& y^{(1)} + \frac{(b-a) y^{(0)}}{b - x^{(0)} y^{(0)} (x^{(0)} + x^{(1)} - y^{(1)})}+
                 \frac{(b-a) x^{(0)}}{a - x^{(0)} y^{(0)} (x^{(0)} + x^{(1)} - y^{(1)})}, \nn \\
  v^{(0)} &=& x^{(0)} + \frac{(b-a) x^{(0)}}{a - x^{(0)} y^{(0)} (x^{(0)} + x^{(1)} - y^{(1)})}, \nn\\
  v^{(1)} &=& x^{(1)} + \frac{(b-a) y^{(0)}}{b - x^{(0)} y^{(0)} (x^{(0)} + x^{(1)} - y^{(1)})}.  \nn
\eea
}\eex

\subsection{The Invariants of ${\bf X}_s$}

Again we denote invariants as in (\ref{xip}) and Figure \ref{xyuv-fig}.
The $4(N-1)$ invariants:
\bea
  && x^{(i)} = \frac{\chi^{(i)}_{m+1,n}}{\chi^{(i+1)}_{m,n}},\quad i = 0,\dots,N-3, \qquad
    x^{(N-2)}  =  \chi^{(N-2)}_{m+1,n} \prod_{j=0}^{N-2} \chi_{m,n}^{(j)}, \nn  \\[-1mm]
 && y^{(i)} = \frac{\chi^{(i)}_{m+1,n+1}}{\chi^{(i+1)}_{m+1,n}},\quad i = 0,\dots,N-3,  \quad
    y^{(N-2)}  =  \chi^{(N-2)}_{m+1,n+1} \prod_{j=0}^{N-2} \chi_{m+1,n}^{(j)}, \nn \\[-1mm]
&&  u^{(i)} = \frac{\chi^{(i)}_{m+1,n+1}}{\chi^{(i+1)}_{m,n+1}},\quad i = 0,\dots,N-3, \quad
    u^{(N-2)}  =   \chi^{(N-2)}_{m+1,n+1} \prod_{j=0}^{N-2} \chi_{m,n+1}^{(j)}, \nn \\[-1mm]
&&  v^{(i)} = \frac{\chi^{(i)}_{m,n+1}}{\chi^{(i+1)}_{m,n}},\quad i = 0,\dots,N-3,  \qquad
    v^{(N-2)}  =  \chi^{(N-2)}_{m,n+1} \prod_{j=0}^{N-2} \chi_{m,n}^{(j)}, \nn
\eea
satisfy $(N-1)$ identities:
\begin{subequations}\label{xyuvs-eq}
\bea
 u^{(i)} v^{(i+1)}  &=& x^{(i+1)} y^{(i)},\quad i = 0,\dots,N-3, \label{xyuvs-eq1}\\
   u^{(N-2)} \prod_{j=0}^{N-2} x^{(j)}  &=& y^{(N-2)} \prod_{j=0}^{N-2}v^{(j)},  \label{xyuvs-eq2}
\eea
and equations (\ref{eq:h-0101-a}) take the form
\bea
  u^{(i)}v^{(i+1)} &=& \frac{(v^{(i)}-1) v^{(i+1)} -(x^{(i)}-1) x^{(i+1)}}{v^{(i)}-x^{(i)}} ,\quad i = 0,\dots,N-3, \label{xyuvs-eq3}\\
    u^{(N-2)} &=&  \left(1+ \frac{1}{x^{(N-2)}-v^{(N-2)}} \left(\frac{a}{X}
                   - \frac{b}{Y} \right)\right)\prod_{j=0}^{N-2} v^{(j)},  \label{xyuvs-eq4}
\eea
\end{subequations}
where $X=\prod_{j=0}^{N-3} (x^{(j)}-1),\; Y=\prod_{j=0}^{N-3} (v^{(j)}-1)$.

\br[The Case $N=2$]
Again, since the lattice equation is just $H1$ in the ABS classification \cite{03-8}, the symmetry ${\bf X}_s$, with $\omega = -1$, leads to the Yang-Baxter map
$$
u = y\left(1 + \frac{a-b}{x-y}\right), \quad v = x\left(1 + \frac{a-b}{x-y}\right),
$$
which is just $F_{IV}$ of the ABS classification of quadrirational maps \cite{04-5}.  Clearly, we may consider this whole family of maps as multi-component generalisations of $F_{IV}$.
\er

\bex[The Case $N=3$]  {\em
In this case, we first define
$$
P_a = a x^{(0)}- (x^{(0)}-1) (y^{(0)}-1) (x^{(0)}x^{(1)}-y^{(1)}),\quad P_b = b x^{(0)}- (x^{(0)}-1) (y^{(0)}-1) (x^{(0)}x^{(1)}-y^{(1)}).
$$
We then have the map
\bea
 u^{(0)} &=& y^{(0)}\left(1- \frac{(a - b) x^{(0)} (y^{(0)}-1)}{(y^{(0)}-1)P_a-y^{(0)} P_b}\right), \nn \\
  u^{(1)} &=& y^{(1)}\left(1- (a-b) \left(\frac{(x^{(0)}-1)y^{(0)}}{P_a}+\frac{(y^{(0)}-1)}{P_b}\right)\right), \nn \\
  v^{(0)} &=& x^{(0)} \left(1-  \frac{(a-b)(x^{(0)}-1)}{P_a}\right), \nn\\
  v^{(1)} &=& x^{(1)}\left(1-  \frac{(a-b)(y^{(0)}-1)x^{(0)}}{P_b}\right).  \nn
\eea
}\eex

\subsection*{Acknowledgements}

PX acknowledges support from the EPSRC grant {\emph{Structure of partial difference equations with continuous symmetries and conservation laws}}, EP/I038675/1, and an {\em Academic Development Fellowship} from the University of Leeds.

%\bibliography{apf}

\end{document}